\documentstyle[aps,prl,multicol,fancyheadings,epsfig,amsbsy,amssymb,amstex]{revtex}

\def\bbbc{{\mathchoice {\setbox0=\hbox{$\displaystyle\rm C$}\hbox{\hbox
to0pt{\kern0.4\wd0\vrule height0.9\ht0\hss}\box0}}
{\setbox0=\hbox{$\textstyle\rm C$}\hbox{\hbox
to0pt{\kern0.4\wd0\vrule height0.9\ht0\hss}\box0}}
{\setbox0=\hbox{$\scriptstyle\rm C$}\hbox{\hbox
to0pt{\kern0.4\wd0\vrule height0.9\ht0\hss}\box0}}
{\setbox0=\hbox{$\scriptscriptstyle\rm C$}\hbox{\hbox
to0pt{\kern0.4\wd0\vrule height0.9\ht0\hss}\box0}}}}
\newcommand{\beq}{\begin{eqnarray}}
\newcommand{\eeq}{\end{eqnarray}}
\newcommand{\s}{\sigma}
\newcommand{\bs}{\bf \sigma}
\newcommand{\bj}{{\bf j}}
\newcommand{\bE}{{\bf E}}
\newcommand{\bH}{{\bf H}}

\newcommand{\br}{{\bf r}}
\newcommand{\on}{{\overline{n}}}

\begin{document}

\title{Spin-drift transport and its applications}
\author{Ivar Martin}
\address{Theoretical Division, Los Alamos National Laboratory,
Los Alamos, NM 87545}

\date{January 25, 2002}

\maketitle

\begin{abstract}

We study the generation of non-equilibrium spin currents in
systems with spatially-inhomogeneous magnetic potentials. For
sufficiently high current densities, the spin polarization can be
transported over distances significantly exceeding the intrinsic
spin-diffusion length. This enables applications that are
impossible within the conventional spin-diffusion regime.
Specifically, we propose dc measurement schemes for the carrier
spin relaxation times, $T_1$ and $T_2$, as well as demonstrate the
possibility of spin species separation by driving current through
a region with an inhomogeneous magnetic potential.

\end{abstract}
\pacs{PACS Numbers: XXXXX}
\begin{multicols}{2}

The rapidly developing field of spin-sensitive electronics
promises to reinvigorate conventional semiconductor
electronics\cite{general}. Utilizing the carriers' spin degrees of
freedom in addition to their charge opens possibilities for
conceptually new {\em spintronic} devices.  Already implemented
are GMR read-heads and non-volatile magnetic RAM (MRAM); there are
also many proposals for other devices \cite{zutic}.  Much of the
promise of spintronic devices lies in the enhanced flexibility of
the spin degrees of freedom compared to electrical charge. While
creating local charge imbalance generally involves large
electrostatic Coulomb energies, non-equilibrium spin polarizations
can be created at a very low cost.  As a consequence, spin
polarizations are inherently susceptible to manipulation, e.g. by
external magnetic fields or by current drive.  Indeed, it has been
observed experimentally that non-equilibrium spin polarization can
be transported over distances exceeding 100 micrometers
\cite{kik}, with only moderate diffusive spread.  This would be
impossible in the case of the non-equilibrium charge
inhomogeneity, since the Coulomb repulsion would blow such
inhomogeneity apart almost instantly.

The immunity of the spin polarization with respect to electrical
screening makes it possible to satisfy  {\em drift} transport
conditions, for which the local spin polarization can be carried
downstream by an electrical current over distances proportional to
the drift velocity. This situation can be obtained for
sufficiently large current densities, when the {\em spin-drift}
length  becomes larger than the {\em spin-diffusion} length (both
defined below). In this Letter we explore this new regime, readily
achievable in spin transport, but almost never realized in
electron-hole transport. It is studied based on general
formulation for spin transport using spin drift-diffusion
equations.

The spin-drift regime is generally favorable for spintronic
purposes as it can provide larger spin-injection depths compared
to the spin-diffusive regime.  Here we propose two specific
applications of the spin-drift regime: (1) dc transport-based
measurement of longitudinal ($T_1$) and transverse ($T_2$) spin
relaxation times; (2) spin species separation by inhomogeneous
magnetic fields. To determine the spin relaxation times, we
propose to measure the spin polarization decay profile in the
presence of constant current; the decay length can then be related
either to $T_1$ or $T_2$, depending on the experimental setup.
Unlike conventional pump-probe techniques, there is no need for
ultrafast time-resolved detection, and the spin polarization
profile can be measured either by optical\cite{kik} or some other
techniques. To spatially separate spin-up from spin-down carriers,
we propose to drive current in a 2D geometry containing a region
of inhomogeneous magnetic field.  This technique can be used as a
flexible alternative to the standard spin injection techniques.

To study spin transport we use the drift-diffusion approximation
for spin currents in the presence of an inhomogeneous
spin-dependent potential $V_\s(\br) = - g \mu_B {\bs} \cdot \bf
H$.  At this point we assume that in the plane of the electron
gas, the applied magnetic field $\bH$ can vary in value but not in
orientation.  Then all carriers in the sample can be divided into
spin-up ($\sigma = 1/2$) and spin-down ($\sigma = -1/2$) species
that feel magnetic potentials equal in magnitude but opposite in
sign, $V_\uparrow(\br) = - V_\downarrow(\br)$.  For the external
potential smaller than the thermal energy in the non-degenerate
case, $V_\s(\br) < k_B T$, or less than the Fermi energy in the
degenerate case, $V_\s(\br) < E_F$, the local carrier density will
remain essentially unchanged, both with and without current drive.
Hence, in what follows we neglect electrical screening effects
since they only modify the results in the order $(V_\s/max(k_B
T,E_F))^2$. In the drift-diffusion approximation, the
spin-polarized current is
\beq\label{eq:j_s}
\bj_\s =  n_\s \mu (q \bE - \nabla V_\s) - q D \nabla n_\s,
\eeq
where $q = \pm e$ is the carrier charge, $n_\s$ the number density
of the spin  species $\s$, $D$ the diffusion  constant, $\mu$ the
mobility, $\bE$ the external electric field, and $\nabla$ denotes
the gradient operator.   We assumed that the transport is unipolar
(e.g. n-doped semiconductor), and the diffusion constant is the
same for both spin species\cite{flatte}. We do not include the
Lorentz force, which also acts on the moving electrons, since it
affects both spin species identically and, moreover, is
compensated through the classical Hall effect for a confined
sample geometry. In practice, local variation of the magnetic
potential can be achieved either through the use of micromagnets
that can create strongly inhomogeneous external magnetic fields,
or by magnetically doping semiconductors, which can produce strong
position dependence in the carrier $g$-factor\cite{ohno}.

The dynamically generated non-equilibrium spin polarization
relaxes towards its equilibrium value.  Under constant drive, the
relaxing spins are replenished by the divergence of the spin
current. In the relaxation-time approximation,
\beq\label{eq:cont}
\nabla\cdot\bj_s &=& - \delta_\s/T_1,\\
\delta_\s &=& n_\s - \overline{n}_\s,\\
\overline{n}_\s &=& n_0 \exp(-V_\s/k_BT)/2.
\eeq
The expression for the equilibrium density, $\overline{n}_\s$,
corresponds to non-degenerate carriers.  For degenerate carriers,
the Boltzman distribution should be replaced by the Fermi-Dirac
distribution.   In the non-stationary case, there is an extra time
derivative, $\partial\delta_\s/\partial t$, on the left hand side
of the continuity equation.  Combining Eqs.~(\ref{eq:j_s}) and
(\ref{eq:cont}), the dc equation for spin density under uniform
drive ($\nabla\cdot \bE = 0$) is
\beq\label{eq:main}
\mu \bE \cdot \nabla(\on_\s + \delta_\s) - \frac{\mu}{q} \nabla
(\delta_\s \nabla V_\s) - D \nabla^2 \delta_\s = - \frac{\delta_\s
}{T_1}
\eeq
The first term on the left hand side represents the drag of the
spin density by the electric field ($\mu \bE = v_d$ -- the drift
velocity), the second term, as we will see, is usually
unimportant, and the last term represents diffusion. Similar
descriptions have been previously considered by several authors
\cite{darryl}, however, with the focus on the {\em diffusive}
regime (neglecting $\bE \delta_\s$ terms).  The drift terms have
only been invoked in the spatially localized regions, such as
depletion layers and Schottky barriers\cite{darryl2}.   The
primary goal of the present work is to explore the {\em
spin-drift} regime that occurs for larger drives, that is when the
spin-drift length, $L_E = \mu E T_1$, exceeds the spin diffusion
length, $L_D = \sqrt{D T_1}$.

We first consider the one dimensional case with spin-polarized
negative half-space.  The static equilibrium density is
\beq
\on_\s - n_0/2  = \s S\Theta(-x),
\eeq
where $\Theta$ is the step function equal to unity for positive
arguments and zero otherwise; $S$ is the static spin density in
the polarized region.  The step in the density profile implies a
step in the magnetic potential, $V_\s$.  The density modification
in the presence of a uniform current drive is described by the 1D
version of Eq.~(\ref{eq:main}). The strongest singularity is
caused by the jump in $V_\s$ at $x = 0$ (second term in
Eq.~(\ref{eq:main})), which implies discontinuity in the density,
$\delta_\s$,
$$\frac{\delta_\s(+0) - \delta_\s(-0)}{\delta_\s(+0) + \delta_\s(-0)} =
\frac{\mu}{2qD}(V_\s(-0) - V_\s(+0)).$$

Within our assumptions about the magnitude of the magnetic
potential, $V_\s < max(k_BT, E_F)$, this correction is small and
hence can be neglected. Also, it causes charge imbalance, and
hence is screened out for distances exceeding the Coulomb
screening length\cite{stu}. The solution of the homogeneous
equation for $\delta_\s$ is
\beq
\delta_\s(x) &=& A\exp(\lambda_1 x) + B\exp(\lambda_2 x), \\
\lambda_{1(2)} &=& \frac{L_E}{2L_D^2}\left( 1 + (-) \sqrt{1 +
4L_D^2/L_E^2}\right).
\eeq
Matching the ($x > 0$) and ($x < 0$) solutions at $x = 0$ and
requiring vanishing $\delta_\s$ at infinity, the spin density,
$s=\delta_\uparrow - \delta_\downarrow$, for the positive drift
velocity ($qE > 0$) is
\beq\label{eq:1D_sol}
s = \frac{S}{\sqrt{1 + 4L_D^2/L_E^2}}
\times
\left\{\begin{array}{cc}
\exp(\lambda_1 x), &x < 0, \\
\exp(\lambda_2 x), &x > 0\end{array}\right. .
\eeq
The solution for the negative drift velocity can be obtained from
this solution as $ -s (-x)$.

Two interesting special cases correspond to the {\em spin-drift}
($L_E \gg L_D$) and the {\em spin-diffusion} ($L_E < L_D$)
regimes. In the spin-drift case,
\beq
s = \left\{\begin{array}{lc}
{S}\exp(-x/L_E), &x> 0, \\
0, &x < 0\end{array}\right..
\eeq
And in the spin-diffusion case,
\beq
s = \frac{S L_E}{2 L_D}\exp(-|x|/L_D).
\eeq

\begin{figure}[htbp]
\begin{center}
\includegraphics[width = 3.0 in]{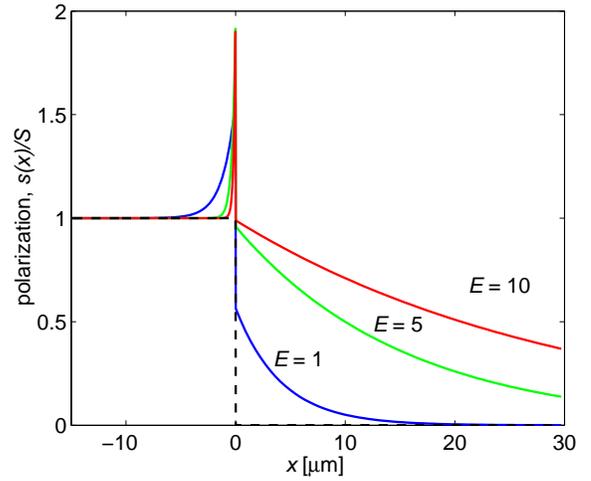}
\vspace{0.5cm} \caption{Spin polarization drag induced by electric
current in 1D geometry, Eq.~(\ref{eq:1D_sol}).  The dashed line
corresponds to the spin polarization profile in the absence of
current. The indicated strength of the electric field, $E$, is in
V/cm.   Note that the stronger the applied electric field is, the
farther away the spin polarization can be dragged. The parameters
are characteristic of $10^{16}$~cm$^{-3}$ Si-doped GaAs: $\mu =
3000$~cm$^2$/Vs, $T_1 = 100$~ns, $T = 1.6$~K.  Carriers are
assumed to be non-degenerate.} \label{fig:1D}
\end{center}
\end{figure}

The series of solutions for various strengths of the electric
field are shown in Figure~\ref{fig:1D}.  The parameters are
characteristic of $10^{16}$~cm$^{-3}$ Si-doped GaAs\cite{kik}:
$\mu = 3000$~cm$^2$/Vs, $T_1 = 100$~ns, $T = 1.6$~K.  The drift
condition for this system is realized for $E > 2\sqrt{k_BT/q\mu
T_1}
\sim 2$~V/cm. Experimentally, much larger electric fields
have been applied to this system\cite{kik}.

Under spin-drift conditions, the characteristic length of the
exponential decay of spin polarization is proportional to the
longitudinal spin relaxation time, $L_E = v_d T_1$.  In this
regime, there is almost one-to-one correspondence between the
distance from the interface and the time spent by electrons in the
region with zero field, $x = v_d t$, which is only weakly altered
by diffusion. Therefore, by measuring the decay length and knowing
the drift velocity, $v_d = \mu E$, one can directly  determine
$T_1$.  An advantage compared to the conventional pump-probe
approach\cite{kik} is that the measurement can be conducted in a
dc setup that doesn't require ultrafast optics and electronics.
The spin polarization detection with sub-micron spatial resolution
can be done by standard optical techniques based on Kerr/Faraday
rotation, or possibly using magnetic resonance force microscopy
(MRFM)\cite{mrfm} by modulating the current at the cantilever
frequency\cite{roukes}.

In addition to the longitudinal relaxation time, $T_1$, which
defines how fast the longitudinal spin polarization disappears, an
almost identical approach can be applied to determine the
transverse spin relaxation time, $T_2$, which measures the
electron spin decoherence time. For that, in the setup described
above, a weak uniform magnetic field, $H_\perp$, should be applied
{\em perpendicular} to the direction of the strong field that
exists at $x < 0$. In the region with the strong field, this
additional component will only weakly alter the the direction of
the total field. However, in the region where originally there was
no magnetic field, this component will cause the precession of the
drifting spin polarization.  The spatial period of the precession
\beq
L_P = 2\pi v_d/\omega_L,
\eeq
is related to the Larmor frequency, $\omega_L = g\mu_B
H_\perp/\hbar$, which is about 62 MHz at $H_\perp = 100$ G for $g
= -0.44 $.  For the driving electric field $E = 10$ V/cm and the
above parameters, $L_P \sim 5\ \mu$m. Such spatial modulation can
be resolved both by optical means or with MRFM.  To determine the
transverse spin relaxation time $T_2$ one needs to measure spatial
dependence of the spin polarization component perpendicular to
$H_\perp$ and then fit it to
\beq\label{eq:T_2}
S \propto \exp(-x/v_d T_2)\cos(2\pi x/L_P).
\eeq
Here we neglected the diffusive spread of the spin polarization,
which will case an analogue of NMR inhomogeneous broadening.  The
diffusive spread will matter, however, only when it becomes
comparable to the precession length, $L_P$. This effectively
places an upper limit on the  distance from  the interface  for
which the precession-induced oscillation can be resolved, $L <
|\mu E L_P^2/D|\sim 250\ \mu$m for the above parameters.  The
upper limit can be modified by the choice of the transverse
magnetic field $B_\perp$ with the only constraint that $\omega_L
T_2 > 1$. Finally, the carrier $g$-factor itself can be determined
from the Larmor frequency, which can be extracted from the
precession length.

\begin{figure}[htbp]
\begin{center}
\includegraphics[width = 3.0 in]{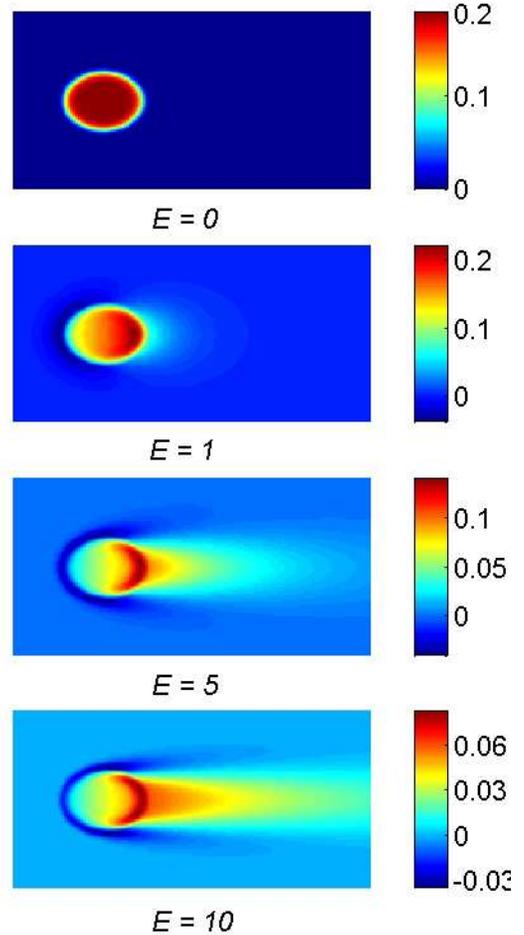}
\vspace{0.5cm} \caption{Spin species separation by inhomogeneous
magnetic field.  An electric current flows in 2D (film) through a
region of localized magnetic field, $V_\s(r) = - 0.6 \s
\exp[-(r/5)^{10}]$, with $r$ in micrometers and energy $V_\s$ in
Kelvin. The spin polarization is measured in the units of the
total density.  The strength of the electric field, $E$, is in
V/cm.  Notice the region of {\it negative} (``minority'') spin
polarization that first forms were the current impinges on the
region of the magnetic field.  For large enough drift velocities,
the ``majority'' and the ``minority'' carriers forms a three-tail
structure that extends over the drift length, $L_E$, downstream.
The material parameters are the same as in Fig.~\ref{fig:1D}.}
\label{fig:2D}
\end{center}
\end{figure}

Another application that is only possible in the spin-drift regime
is the {\em spatial} separation of spin species by driving
electric current through a region with an inhomogeneous magnetic
field. To demonstrate this effect, we consider a two-dimensional
system with a compact region containing magnetic field.  Unlike
the one-dimensional case considered above, the ``minority''
(opposite to magnetic field) spin carriers can avoid this region
by flowing around it. Indeed, this is what we observe by solving
numerically the 2D spin-relaxation drift-diffusion equations,
Fig.~\ref{fig:2D}.  The calculation is performed on a $100 \times
150$ lattice that spans $30\times45\ \mu$m region for parameters
relevant to Si-doped GaAs (specified above). The field is
localized in 10 $\mu$m disk. For large enough electric field,
there is formation of the negative spin polarization where the
current impinges on the boundary of the magnetic field
region\cite{dassarmaPN}. The minority spin polarization is then
dragged downstream by the current until it relaxes and diffuses
away.   Notice that in the zero-current state there is $no$
negative spin polarization in the system.  The ``majority'' spin
polarization is ripped off the local field region, and together
with the ``minority'' carriers forms a three-tail structure that
extends over the spin-drift length, $L_E$, downstream. Therefore,
driving current through this system leads to an effective
separation of the two spin species.  A similar effect can be
expected also in the ballistic regime realized in the ultra-high
mobility 2DEG structures.

The effect that we discussed above is analogous to the classic
Stern-Gerlach effect; however, unlike the original effect the spin
separation can be achieved even with electrons.  In the original
Stern-Gerlach experiment,  {\em neutral} spin-1/2 particles
passing through a region with a magnetic field  gradient get
deflected by the force $F = g \mu_B (\s\cdot\nabla)\bH$, and
produce two spots on the screen that represent two possible
projections of spin $\s$.  For charged particles, the situation is
complicated due to the Lorentz force that acts to scramble the
separation.  It was first pointed out by Bohr that in the case of
electrons, the observation of Stern-Gerlach effect is impossible
because the deflection by the Lorentz force exceeds the
spin-related splitting\cite{SGrec}.  The Bohr argument does not
apply, however, to constrained geometries, e.g. a thin film
sample, where the Lorentz force is compensated by the Hall
voltage.  This makes the {\em solid-state Stern-Gerlach} effect
possible.  It is important to stress, however, that despite the
appearance, the mechanism for spin separation here is distinctly
different from the the classic Stern-Gerlach effect.  A {\em
time-dependent} Stern-Gerlach effect was recently considered by
Fabian and Das Sarma\cite{dassarmaSG}.

In the above discussion we assumed a possibility of a magnetic
field profile that is inhomogeneous but always pointing along the
same line, e.g. perpendicular to the sample plane.  This is an
important assumption because the fringe fields have ability to
destroy spin polarization through precession.  For the $10^{16}$
Si:GaAs, a 100 Gauss fringe field at 10 V/cm drive could scramble
the polarization after 5 $\mu$m.  However, since the fringe fields
decay rapidly in space, their detrimental effects can be reduced
by increasing the drive or by reducing the size of the
micromagnets.  Moreover, for thin-film samples, the effects of the
fringe fields can be essentially eliminated by placing the sample
between two closely spaced magnet terminals (split ring) or
between two ``identical'' magnets.  Alternatively, one could
selectively boost the $g$-factor, e.g. by locally doping II-VI
semiconductors with Mn. This way, when placed even in a relatively
weak uniform external magnetic field, different parts of the
sample will feel very different magnetic potentials.  By combining
these approaches, one can engineer a variety of strong magnetic
potentials which are not affected by the fringe fields.

In summary, we explored the spin-drift regime, which is realized
for sufficiently large but realistic current densities in
semiconductors.  In this regime, the spin density can be carried
by the current over distances controlled by the spin-drift length,
which can  significantly exceed the spin-diffusion length.  This
opens a possibility of designing spintronic devices with the spin
properties dynamically controllable by the applied bias. This is
in sharp contrast with the conventional particle-hole
semiconductor electronics where electrostatic screening
practically precludes large scale dynamic manipulation of the
relative particle-hole density. As specific applications, we
proposed measurement schemes for the carrier spin relaxation
times, $T_1$ and $T_2$, as well as demonstrated the possibility of
spin species separation by driving current through a region with
an inhomogeneous magnetic potential.  The latter is analogous to
the classic Stern-Gerlach effect.

Upon completion of this paper, author became aware of the related
one-dimensional study of the spin-drift regime by Yu and
Flatt\'{e}\cite{yu}.

We thank S.A. Crooker, J. Albrecht, D.L. Smith, J. Fabian, and
S.A. Trugman for useful discussions.  This work was supported by
DARPA SPINs program.

\end{multicols}
\end{document}